# Technical Comment
# The impact of population-wide rapid antigen testing on SARS-CoV-2 prevalence in Slovakia

Matúš Medo[1,*], Martin Šuster[2,*], Katarína Boďová[3], Alexandra Bražinová[4], Broňa Brejová[3], Richard Kollár[3], Vladimír Leksa[5,8], Jana Lindbloom[6], Jozef Nosek[7], Veda pomáha COVID-19[#], Tomáš Vinař[3]

[1] Department of Radiation Oncology, Inselspital, University Hospital of Bern, and University of Bern, 3010 Bern, Switzerland
[2] National Bank of Slovakia, Imricha Karvaša 1, Bratislava, Slovakia
[3] Faculty of Mathematics, Physics and Informatics, Comenius University in Bratislava, Mlynská dolina, 842 48 Bratislava, Slovakia
[4] Institute of Epidemiology, Faculty of Medicine, Comenius University in Bratislava, Špitálska 24, 813 72 Bratislava, Slovakia
[5] Institute for Hygiene and Applied Immunology, Center for Pathophysiology, Infectiology and Immunology, Medical University of Vienna, Lazarettgasse 19, A-1090, Vienna, Austria
[6] Institute for Sociology, Slovak Academy of Sciences, Klemensova 19, 813 64 Bratislava, Slovakia
[7] Faculty of Natural Sciences, Comenius University in Bratislava, Ilkovičova 6, 842 15 Bratislava, Slovakia
[8] Laboratory of Molecular Immunology, Institute of Molecular Biology, Slovak Academy of Sciences, Dúbravská cesta 21, 845 51, Bratislava, Slovakia

[*] The first two authors contributed equally to this work
[#] For a complete list of the contributing member of Veda pomáha COVID-19 initiative, see the Acknowledgement section

**Abstract**

Pavelka et al. (*Science*, *Reports, 7 May 2021*) claim that a single round of population-wide antigen testing in Slovakia reduced the observed COVID-19 prevalence by 58%, and that it played a substantial role in curbing the pandemic. We argue that this estimate, which is based on incorrect assumptions, is exaggerated, and that the relief was short-lived with little effect on mitigating the pandemic.

Pavelka et al.[1] analyze the effects of non-pharmaceutical interventions (NPIs) implemented in Slovakia in the fall of 2020 (see Fig.1 for the timeline) as a response to a worsening COVID-19 epidemiological situation[2]. They estimate that one round of population-wide antigen testing coupled with standard NPIs lowered the (observed) prevalence by 58%, and they extrapolate that repeated mass testing decreases the prevalence by the same factor in each round. We propose that: (i) effects of the mass testing were much lower and temporary; (ii) their repeatability is questionable; and (iii) other externalities should be considered to evaluate long-term effects.

Some problems stem from an incorrect use of terminology. Pavelka et al. define "*observed infection prevalence*" as the proportion of positive results in mass antigen testing. They use this term interchangeably with the term "prevalence" (proportion of infected individuals in the population[3]), even in the study title and abstract.

The estimate of a 58% decrease in observed prevalence in counties with multiple rounds of mass testing compares the test positivity on two successive weekends. However, the two population samples are influenced by multiple asymmetrical biases. First, individuals positive in the first round of testing (T1) and their whole households were instructed to quarantine and not to participate in the second round (T2), which caused a significant underestimation of prevalence in T2. This effect is also seen in the microsimulation model of Pavelka et al., where prevalence and test positivity are similar in the pilot round (T0), but they differ markedly in subsequent rounds (Fig 2a). A similar discrepancy is also shown in Fig.3b of Pavelka et al. (green line vs. symbols). Second, self-selected participants were motivated by lockdown exemptions, which were a necessary inducement to participate[4]. Experience from T1 possibly decreased participation in T2 for individuals suspecting their own positivity to avoid loss of income from a quarantine of the entire household. Third, many residents of counties exempted from T2 needed a negative test for commuting and were tested in other participating locations, where they influenced the positivity[5].

Moreover, the authors consider the value of 58% a "*robust*" estimate (95% confidence interval 57-58%), even though they acknowledge their inability to explain a significant variance among counties, which ranged from 47% to 70% in the counties with at least 50 thousand participants.

Pavelka et al. use mass antigen testing both as an intervention and as its efficacy measurement, which is often a source of unforeseen biases[6]. The fact that T2 was only implemented in selected counties provides a quasi-experimental setup to estimate mass testing efficacy. The RT-qPCR incidence (measured independently from the intervention) was reduced by 40% in counties with T2 between the week before T2 and the second week after, compared to a reduction by 22% in counties without T2 (Fig.1a) – a 23% reduction attributable to the mass testing. Kahanec et al.[7] corrected for biases stemming from differences in the effectiveness of antigen testing in low- and high-prevalence areas, estimating 25-30% reduction over the same period.

This exaggeration of prevalence decrease is independently confirmed by other data (Fig.1b): hospital admissions (10-day lag after infection[8]) decreased by about 30% from their peak (hardly the "*sharp decrease in new admissions*"[1]), and excess deaths (15-20 day lag[9]) decreased by about 24%, both consistent with a prevalence reduction of 20-30%. In both cases, the temporary two-week decline was followed by a long-term steady increase.

Pavelka et al. use a microsimulation to model prevalence in simulated populations under different parameter settings. They assume 100% test sensitivity, in stark contrast with the well-documented lower sensitivity of antigen tests[10]. Moreover, no simulation scenario takes into account the increase of mobility after T1 (Fig.1c). Changing these assumptions alone dramatically reduces the cumulative effects of mass testing from 90% to 55% over three rounds (Fig.2b). The strong effects presented in Fig.3 of Pavelka et al. depend on unrealistic assumptions about test sensitivity and social implications (ignoring a false sense of security after negative tests).

Several claims of Pavelka et al. are based on epidemic growth of 4.4% per day at the time of T1, estimated from the RT-qPCR incidence. Regrettably, the details of the computation of this number, or of the 70% decrease in prevalence compared to unmitigated growth, are not provided by the authors. However, a nationwide improvement had been observed already before T1 (Fig.1a,b,d), suggesting a much lower epidemic growth (or even decline) at the time, which makes the basis of this computation unfounded.

Finally, Pavelka et al. conclude that mass testing likely had *"a substantial effect in curbing the pandemic in Slovakia"*. Yet, the effects were short-lived. Mobility increased steadily starting from T1, indicating changes in social behavior (Fig.1c). The long-term increase of RT-qPCR incidence began during T2, the positivity of routine antigen tests and the average viral load in positive RT-qPCR tests began to increase several days later (Fig.1d). Hospital admissions started to rise two weeks after T2, followed by a steady increase in excess deaths (Fig.1b), eventually resulting in overloaded hospitals, a 4-month lockdown from January 2021, and the world's foremost position in reported deaths per capita[11] in February 2021.

The first mass antigen testing in Slovakia was organized under unique circumstances difficult to repeat on the same scale. It was facilitated by an extensive involvement of the army and municipalities, minute-to-minute media coverage, the commitment of numerous medical personnel and volunteers (including international aid), and social effects of participants being tested simultaneously. Spontaneously emerging social pressure likely increased not only testing attendance, but also observance of the quarantine. However, all this led to only a temporary decrease in incidence, much lower than estimated by Pavelka et al. Such a mobilization cannot be conceived as a viable long-term strategy, which has been proven in Slovakia itself, as subsequent population-wide antigen testing in 2021 failed to produce anticipated results.

Based on these arguments, it is questionable whether mass testing in Slovakia should serve as a model for other countries.

**Acknowledgements.** The authors would like to thank Michael Z. Lin and Marc Bonten for consultation. The following members of **Veda pomáha COVID-19** initiative contributed to the article: Eduard Baumohl[1], Fedor Blaščák, Branislav Bleha[2], Vladimír Boža[3], Peter Celec[4], Jaroslav Frinda[5], Damas Gruska[3], Július Hodosy[4], Peter Jacko[6], Daniel Kerekes[7], Boris Klempa[8], Juraj Kopáček[8], Radomír Masaryk[9], Silvia Pastoreková[8], Jarmila Pekarčíková[10], Peter Sabaka[11], Stanislava Segečová[12], Martin Smatana, Peter Szolcsányi[13], Ľubomír Tomáška[2], Michal Vašečka[14], Peter Visolajský[15], Kristína Visolajská, Peter Zmeko[16]

[1]Faculty of Commerce, University of Economics in Bratislava, Bratislava, Slovakia
[2]Faculty of Natural Sciences, Comenius University in Bratislava, Bratislava, Slovakia
[3]Faculty of Mathematics, Physics and Informatics, Comenius University in Bratislava, Bratislava, Slovakia
[4]Institute of Molecular Biomedicine, Faculty of Medicine, Comenius University in Bratislava, Bratislava, Slovakia
[5]Faculty of Operation and Economics of Transport, University of Zilina, Žilina, Slovakia
[6]Management School, Lancaster University, United Kingdom

[7]Faculty of International Relations, University of Economics in Bratislava, Bratislava, Slovakia

[8]Institute of Virology, Biomedical Research Center of the Slovak Academy of Sciences, Bratislava, Slovakia

[9]Faculty of Social and Economic Sciences, Comenius University in Bratislava, Bratislava, Slovakia

[10]Faculty of Health Care and Social Work, Trnava University, Trnava, Slovakia

[11]Faculty of Medicine, Comenius University, Bratislava, Slovakia

[12]Specialized Hospital of Saint Svorad, Zobor, Nitra, Slovakia

[13]Department of Organic Chemistry, Slovak University of Technology, Bratislava, Slovakia

[14]Faculty of Media, Pan-European University, Bratislava, Slovakia

[15]Faculty Hospital Nitra, Nitra, Slovakia

[16]Faculty of Law, Comenius University in Bratislava, Bratislava, Slovakia

**Author's contributions.** TV, MS, RK conceived the study. MM, MS, RK, KB collected and analyzed the data. All authors participated in discussions, writing, and proofreading of the manuscript. All authors approved the final version of the manuscript.

**Funding.** No funding is declared.

**Conflict of interest.** No conflict of interest is declared.

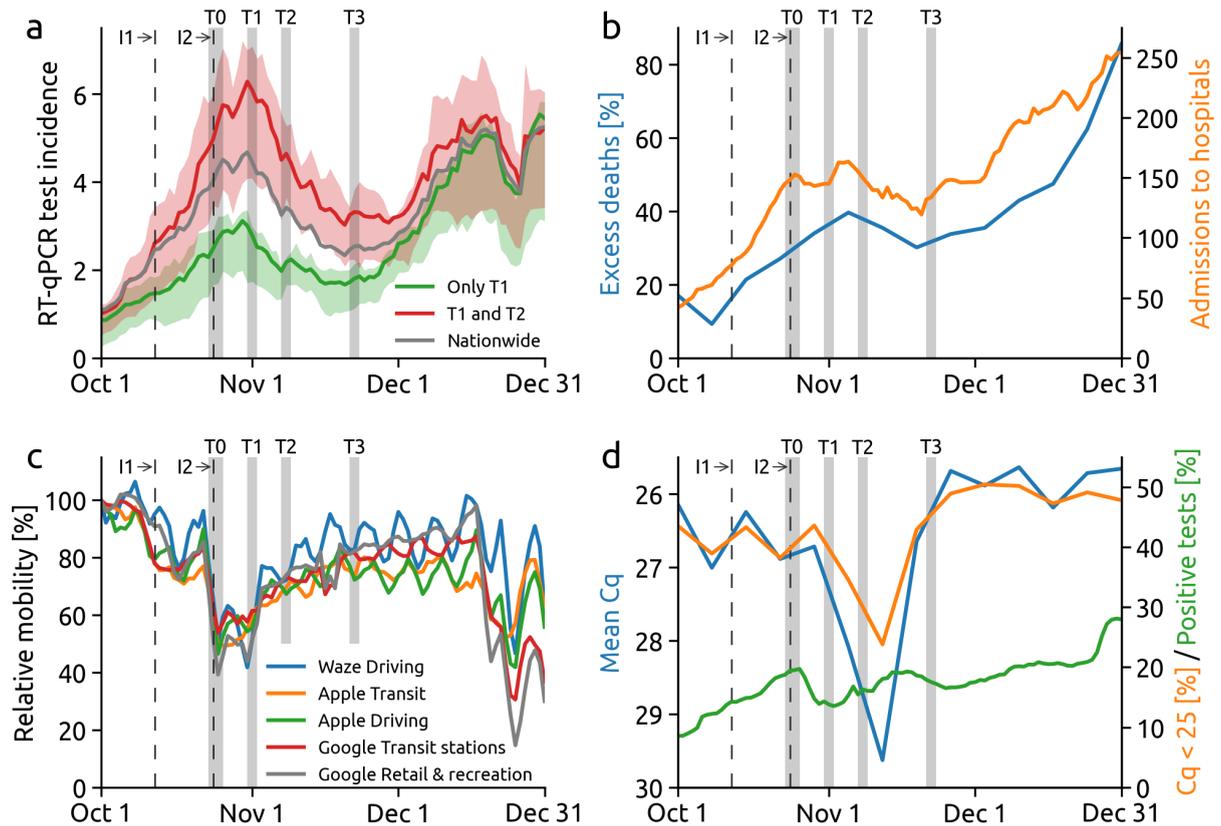

**Figure 1. Temporal evolution of epidemiological indicators in Slovakia between October 1 and December 31, 2020 and their relation to non-pharmaceutical interventions (NPIs).** Timeline of NPIs introduced between October 12 and November 8. I1 (October 12): limiting social contacts (closures of schools, restaurants, cultural venues, sport and recreational facilities, churches, etc., mandatory masks outdoors); I2 (October 24): nation-wide "stay-at-home" order; T0 (October 23-25): pilot mass testing in four counties; T1 (October 31-November 1): nation-wide first round of mass testing; T2 (November 7-8): second round of mass testing in 45 counties (out of 80) with positivity at least 0.7%; T3 (November 21-22): third round of mass testing in 447 municipalities with positivity at least 1% in T1 or T2. For all indicators, moving averages centered in the middle are shown to smooth out day-to-day variation. For mobility data, a 3-day window is used, all other indicators use a 7-day window. **(a) RT-qPCR test incidence per 10,000 people.** Total incidence for all counties (black), incidence for counties with both T1 and T2, but without T0 (red), and for countries with T1 only (green) is shown. Nation-wide incidence started to decline two days before T1. **(b) Hospital admissions and excess deaths.** Hospital admissions declined even before T1 (from October 25 to November 2). After T1, both indicators declined for a short period, resuming a sharp upward trend by the end of November. **(c) Mobility relative to October 1, 2020.** Data from Google mobility (retail, stations), Apple (driving, transit), and Waze suggest a systematic increase of mobility in November 2020. **(d) Positivity of RT-qPCR tests and quantification cycle (Cq) values.** Cq values were reported only on a portion of the tests which were performed by Unilabs diagnostic laboratories. The average viral load and the proportion of the RT-qPCR tests with high viral loads (Cq<25) both started to decrease in October before T1. Lower viral loads are indicative of lower $R_e$[12].

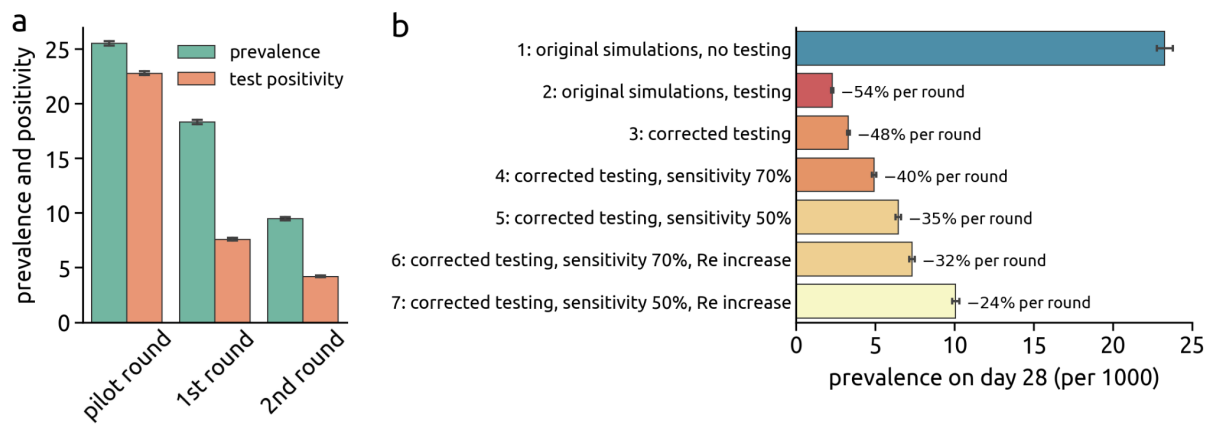

**Figure 2. Impacts of methodology and model parameters on prevalence reduction in the microsimulation by Pavelka et al.** The microsimulation presented in Fig.3 of Pavelka et al. based on the situation in a representative county with the pilot round and two additional rounds of mass testing. All values presented here were obtained with the code and parameters provided by Pavelka et al. in their supplementary material. **(a) Comparison of disease prevalence and antigen test positivity (observed prevalence according to Pavelka et al.).** The results correspond to the scenario with three rounds of testing with extended measures (Re=1) with full household compliance as presented in Fig.3b of Pavelka et al. In the pilot round, test positivity is a good estimate of the disease prevalence. In the following rounds, systematic biases, such as quarantine of infected household members, result in testing positivity significantly underestimating the disease prevalence. This leads to a systematic overestimation of NPI effects. In particular, test positivity shows 67% reduction between the pilot round and the 1st round, while the real prevalence only decreases by 28%. **(b) Prevalence two weeks after the second round under various scenarios.** The microsimulation model is very sensitive to parameters that in Pavelka et al. simulations were set as unreasonably favorable for mass antigen testing. Bar 1: Prevalence in the base scenario with extended measures (Re=1) and no mass testing as presented in Fig.3b of Pavelka et al. Bar 2: Original simulation, mass testing with 100% test sensitivity, full household compliance led to the average reduction of prevalence by 54% per simulation round. However, the code of Pavelka et al. mistakenly performed testing on individuals below 10 and above 65 years, thus mistakenly performing approximately 37% more tests compared to the model county. Bar 3: Simulation results after correcting this issue. Bar 4: Changing test sensitivity to 70% instead of 100%, a value considered as realistic by Pavelka et al. Bar 5: Changing test sensitivity to 50%. Bars 6,7: People were motivated to participate in the mass testing by relief from extended measures as demonstrated by the increase in mobility in Figure 1A. By adding gradual growth of Re from 1 before the pilot round to 1.15 one week after, the effect of mass testing is significantly reduced again.